\documentclass[twocolumn,preprintnumbers,prd,longbibliography,nofootinbib]{revtex4-1}

\usepackage{graphicx}
\usepackage{dcolumn}
\usepackage{bm}
\usepackage{amstext}
\usepackage{amssymb}
\usepackage{amsmath}
\usepackage{graphicx}
\usepackage{color}
\usepackage{delimset}

\def\dd{\delta\!\!\!{}^-\!}

\usepackage{tikz}
\usetikzlibrary{decorations.pathmorphing}
\usetikzlibrary{decorations.markings}
\usetikzlibrary{positioning, shapes, snakes, arrows}

\tikzset{
	graviton/.style={line width=.8pt, -latex,decorate, decoration={snake, segment length=4pt,amplitude=1.8pt, pre length=.1cm, post length=.25cm}},
	worldline/.style={gray, line width=1pt},
	worldlineBold/.style={black, line width=.6pt},
	zUndirected/.style={line width=1pt},
	zParticle/.style={line width=1pt,postaction={decorate},decoration={markings,mark=at position .6 with {\arrow[#1]{latex}}}},
	zParticleF/.style={line width=1pt,postaction={decorate}},
	cscalar/.style={line width=1pt,postaction={decorate},decoration={markings,mark=at position .6 with {\arrow[#1]{latex}}}},
	cscalar2/.style={line width=1pt,postaction={decorate},decoration={markings,mark=at position .8 with {\arrow[#1]{latex}}}},
	photon/.style={line width =.8pt, decorate, decoration={snake, segment length=4pt, amplitude=1.8pt,  pre length=.1cm, post length=.1cm}}
}

\def\dd{\delta\!\!\!{}^-\!}
\xdefinecolor{refcolor}{rgb}{0.5,0,0}
\usepackage[colorlinks=true, filecolor=refcolor, citecolor=refcolor, linkcolor=refcolor, urlcolor=refcolor, menucolor=refcolor]{hyperref}

\makeatletter
\newlength{\apb@width}
\newcommand{\autoparbox}[2][c]{\settowidth{\apb@width}{#2}\parbox[#1]{\apb@width}{#2}}

\makeatother

\makeatletter
\def\mr@ignsp#1 {\ifx\:#1\@empty\else #1\expandafter\mr@ignsp\fi}%
\newcommand{\multiref}[1]{\begingroup
\xdef\mr@no@sparg{\expandafter\mr@ignsp#1 \: }%
\def\mr@comma{}%
\@for\mr@refs:=\mr@no@sparg\do{\mr@comma\def\mr@comma{,}\ref{\mr@refs}}%
\endgroup}
\makeatother

\begin{document}
\preprint{
HU-EP-22/20
}

\title{Binary Dynamics from Worldline QFT for Scalar-QED}
\author{Tianheng Wang}
\email{wangtianheng@itp.ac.cn}

\affiliation{%
Institute of Theoretical Physics, Chinese Academy of Sciences,
55 Zhongguancun Road East, 100190 Haidian District, Beijing, China
}
\affiliation{%
Institut f\"ur Physik und IRIS Adlershof, Humboldt-Universi\"at zu Berlin,
Zum Gro{\ss}en Windkanal 6, 12489 Berlin, Germany
}

\date{\today}

\begin{abstract}
We investigate the worldline quantum field theory (WQFT) formalism for scalar-QED and observe that a generating function emerges from WQFT, from which the scattering angle ensues. This generating function bears important similarities with the radial action in that it requires no consideration of exponentiation of lower-order contributions. We demonstrate the computations of this generating function and the resulting scattering angle of a binary system coupled to electromagnetic field up to the third order in the Post-Minkowskian expansion (3PM).
\end{abstract}

\maketitle

Recent developments in gravitational-wave physics~\cite{LIGOScientific:2016aoc,LIGOScientific:2017vwq,LIGOScientific:2018mvr,LIGOScientific:2020ibl,LIGOScientific:2021usb} call for innovations of theoretical framework that facilitate both numerical~\cite{Pretorius:2005gq,Campanelli:2005dd,Baker:2005vv} and analytical~\cite{Buonanno:1998gg,Goldberger:2004jt,Kol:2007rx,Gilmore:2008gq,Foffa:2011ub,Foffa:2016rgu,Porto:2017dgs,Blumlein:2019zku,Foffa:2019rdf,Foffa:2019yfl,Blumlein:2020pog,Blumlein:2020znm,Bini:2020nsb,Bini:2020hmy,Blumlein:2021txe,Blumlein:2020pyo,Foffa:2020nqe,Blumlein:2021txj,Kalin:2020fhe,Kalin:2020lmz,Mougiakakos:2021ckm,Riva:2021vnj,Dlapa:2021npj,Dlapa:2021vgp,Goldberger:2020fot,Kim:2021rfj,Cho:2022syn,Kalin:2020mvi} high-precision computations of the dynamics of binary black hole or neutron star mergers. 

It has proven fruitful to extract classical observables from scattering amplitudes in perturbative quantum field theories~\cite{Bjerrum-Bohr:2018xdl,Cheung:2018wkq,Bern:2019nnu,Bern:2019crd,Neill:2013wsa,Cristofoli:2021vyo}, thanks to modern tools based on on-shell techniques~\cite{Bern:1994zx,Bern:1994cg,Britto:2004nc,Bjerrum-Bohr:2013bxa,Luna:2017dtq,Kawai:1985xq,Bern:2008qj,Bern:2010ue,Bern:2012uf,Bern:2019prr} and effective field theory~\cite{Cheung:2018wkq,Bjerrum-Bohr:2002gqz}. However, to expose the classical quantity, amplitudes-based approaches often requires a delicate analysis which removes quantum and superclassical contributions alike~\cite{Bern:2019nnu,Bern:2019crd,Bern:2020buy,Bern:2021dqo}. 
Alternative methods that capture classical observables more directly are therefore in demand and several explorations in this direction~\cite{Kosower:2018adc,Maybee:2019jus,Brandhuber:2021kpo,Brandhuber:2021eyq} have been shown to be beneficial.

It is in this light that the worldline quantum field theory (WQFT)~\cite{Mogull:2020sak}, in which worldline degrees of freedom are quantised, is formulated, providing a formal link between black hole observables extracted from scattering amplitudes and time-ordered correlators in WQFT. WQFT Feynman rules circumvent the need for the effective potential in traditional worldline EFT methods~\cite{Goldberger:2004jt,Goldberger:2006bd,Kol:2007bc} and streamline loop calculations encountered in amplitudes-based approaches to summing over diagrams of tree topologies only, yielding classical observables directly. Recent applications of WQFT involve a series of work on spinning black holes ~\cite{Jakobsen:2021smu,Jakobsen:2021lvp,Jakobsen:2021zvh} and the state-of-the-arts derivations of the conservative momentum impulse and the spin kick up to the third order in Post-Minkowskian (3PM) expansion and quardratic order in spin have been obtained from WQFT~\cite{Jakobsen:2022fcj}.

As established in amplitudes-based approaches, conservative and radiative dynamics in classical relativistic scattering can be extracted from the eikonal phase~\cite{Parra-Martinez:2020dzs,Bern:2021dqo,DiVecchia:2021bdo,Heissenberg:2021tzo,DiVecchia:2022nna}. Inspired by the eikonal approximation, an amplitude-action relation has been revealed~\cite{Bern:2021dqo} and the \emph{radial action}~\cite{Damgaard:2021ipf,Kol:2021jjc,Bjerrum-Bohr:2021wwt,Bern:2021xze}  
serves as another generating function for the scattering angle. 
Another closely related generating function is defined in the heavy-particle EFT~\cite{Brandhuber:2021eyq} which agrees with the radial action in their real parts, but differs in the imaginary part. 
One crucial difference between these functions and the standard eikonal exponentiation is that iterations from lower orders can be discarded for the former.

WQFT is expected to have the potential of capturing such generating functions too. The classical eikonal phase can be obtained from WQFT in various contexts up to 2PM/next-to-leading order (NLO)~\cite{Mogull:2020sak,Jakobsen:2021zvh,Shi:2021qsb,Bastianelli:2021nbs}. However, the calculation of the eikonal phase at 3PM and beyond in WQFT remains somewhat ambiguous in the $i\epsilon$-prescription of the worldline propagator.
On the other hand, it is conceivable that WQFT speaks more directly to a generating function whose classical part is readily isolated than the eikonal. 
In this letter, we seek to explore the construction of such a generating function from WQFT.

In this letter, we consider the WQFT counterpart of scalar-QED as a toy model, which is shown to be a useful playground for higher PM gravitational computations~\cite{Westpfahl:1985tsl,Buonanno:2000qq,Bern:2021xze}. We illustrate that a generating function emerges from WQFT in a highly streamlined fashion, which reproduces both the conservative and radiative contributions of the scattering angle. This generating function bears similarities with the radial action and the eikonal exponentiation. The WQFT integrands can be made to match with those in the heavy-mass limit of scalar-QED in the comparable-masses sector and in those diagrams responsible for the radiation reaction. We expect these observations to carry over straightforwardly to WQFT in gravitational background.

\paragraph*{WQFT Formalism for Scalar-QED}
The worldline action describing a charged massive non-spinning point-particle in an electromagnetic background reads~\cite{Schubert:2001he,Edwards:2019eby} 
\begin{align}
\mathcal S_i =  -m_i\int \mathsf{d}\sigma \left[ {1\over2}  (\eta^{-1}\dot{x}_i^2+\eta) + ie {q_i\over m_i} A_\mu\dot{x}^\mu_i\right]\,,
\end{align}
where the worldline coordinate $x^\mu$ is parameterised by $\sigma$ and $\dot{x}^\mu = \mathsf{d}x^\mu / \mathsf{d} \sigma$. $q_i$ and $m_i$ denote the charge and mass of the scalar $i=1,2$. The worldline is coupled to the electromagnetic (EM) field $A_\mu$ and the bulk theory is simply given by the usual EM action. For convenience, we set the einbein $\eta(\sigma)=1$. 
As shown in~\cite{Mogull:2020sak}, specialising the photon to plane waves of fixed momenta and polarisations, the photon-dressed Feynman-Schwinger propagator~\cite{Feynman:1950ir} can be identified with the path integral for the WQFT correlator, with external legs amputated through the LSZ reduction. 

Expanding the worldline around straightline trajectories $x^\mu_i = b_i^\mu + u_i^\mu \sigma + z_i^\mu(\sigma)$, the WQFT Feynman rules are readily expressed in frequency/momentum space: $z_i^\mu(\sigma) = \int_\omega e^{-i\sigma\omega} z_i^\mu(\sigma)$ and $A_\mu(x) = \int_k e^{-ik\cdot x} A_\mu(-k)$, where we have used the shorthand notations $\int_{\omega/ k}$ as introduced in~\cite{Mogull:2020sak}. The explicit expressions of WQFT Feynman rules for worldine-photon interactions are given in Appendix \ref{app:FeynRules}.

Inspired by the eikonal exponentiation~\cite{Mogull:2020sak}, we consider the phase identified with the WQFT path integral in the classical limit,
\begin{align}
e^{i\delta}=\mathcal Z_{\text{WQFT}} = \int \mathcal D [A ] \prod_{j=1}^2\mathcal D [ z_j ]\, e^{i \left(\mathcal S_{\text{EM}}+\sum_{j=1}^2 \mathcal S_j \right) }\,,
\end{align}
where $\mathcal S_{\text{EM}}$ denotes the standard action for the electromagnetic field in the bulk. We note that the identification above is designed to hold in the classical limit. Hence the phase $\delta$ is a purely classical quantity. That is $\delta$ is uniform in $\hbar$ and only admits an expansion in the coupling constant $e^2$. 
Taking logarithm on both sides, we identify $\delta$ at each order of $e^2$ with the sum of connected WQFT diagrams, without iteration corrections from lower orders, which sets it apart from the eikonal approach proposed in~\cite{Mogull:2020sak}. It may be tempting to identify it with the radial action due to the similar definitions; but preliminary evidences suggest that differences occur in their respective imaginary parts. Similar to the HEFT phase~\cite{Brandhuber:2021eyq}, we restrict ourselves to the real parts of this generating function and the resulting scattering angle. The imaginary part is beyond the scope of this letter.

The evaluation of WQFT path integrals is normally sensitive to the $i\epsilon$-prescription of the worldline propagator. 
We observe that only the \emph{principal-value} part of the time-symmetric propagator~\cite{Mogull:2020sak} is relevant for the construction of this generating function. Hence we propose the \emph{principal-value} prescription for the worldline propagator and the propagator simply reads
\begin{align}
\begin{tikzpicture}[baseline={(current bounding box.center)}]
\coordinate (in) at (-1,0);
\coordinate (out) at (1,0);
\coordinate (x1) at (-0.5,0);
\coordinate (x2) at (0.5,0);
\node (z1) at (-0.5, -0.3) {$z^\mu$};
\node (z2) at (0.5, -0.3) {$z^\nu$};
\node (w) at (0,0.3) {$\omega$};
\draw [dotted] (in) -- (x1);
\draw [zUndirected] (x1) -- (x2);
\draw [dotted] (x2) -- (out);
\draw [fill] (x1) circle (.06);
\draw [fill] (x2) circle (.06);
\end{tikzpicture} =
-{i\over m} {\eta^{\mu\nu} \over \omega^2}\,.
\end{align}
This treatment is reminiscent of~\cite{Shi:2021qsb,Brandhuber:2021eyq,Bjerrum-Bohr:2021wwt,Bern:2021yeh}.

The Feynman rules are given in terms of kinematic variables $b_i^\mu$ and $u_i^\mu$, the interpretation of which depends on the worldline trajectory they describe~\cite{Mogull:2020sak}. The kinematics of the $2\rightarrow2$ scattering is given by the momenta: 
\begin{align}
p_1 = \bar{p}_1 +q/2\,,&~~~~~&p_2=\bar{p}_2-q/2\,,\nonumber\\
p'_1 = \bar{p}_1-q/2\,, &~~~~~&p'_2=\bar{p}_2+q/2\,,\nonumber
\end{align}
with $p_i^2 = p^{\prime 2}_i =m_i^2$ and $\bar{p}_i^2 = \bar{m}_i^2$. The initial trajectory ($\sigma=-\infty$) corresponds to $p_i^\mu = m_i u_i^\mu$ and the initial impact parameter is given by $b^\mu=b_1^\mu-b_2^\mu$. The in-between trajectory ($\sigma=0$) corresponds to the ``barred variable'' $\bar{p}^\mu_i = \bar{m_i} \bar{u}^\mu_i$ and $\bar{b}^\mu$. The differences between the two sets of variables come at $\mathcal O(\mathsf{q}^2)$. Similar to the observations in~\cite{Brandhuber:2021eyq}, the phase $\delta$ is free from iterations and hence the barred variables can be traded with the unbarred ones at no cost.

\paragraph*{1PM \& 2PM}
At the leading (1PM) and subleading (2PM) orders, the phase $\delta$ is given by
\begin{align}\label{eq:delta0delta1}
&i\left(\delta^{(0)} +\delta^{(1)} \right) \nonumber\\
=& \begin{tikzpicture}[baseline={(current bounding box.center)}]
\coordinate (in1) at (-1,0);
\coordinate (out1) at (1,0);
\coordinate (in2) at (-1,-1);
\coordinate (out2) at (1,-1);
\coordinate (x1) at (0,0);
\coordinate (x2) at (0,-1);
\draw [dotted] (in1) -- (x1);
\draw [dotted] (x1) -- (out1);
\draw [dotted] (in2) -- (x2);
\draw [dotted] (x2) -- (out2);
\draw [photon] (x1) -- (x2);
\draw [fill] (x1) circle (.06);
\draw [fill] (x2) circle (.06);
\end{tikzpicture} 
+
\begin{tikzpicture}[baseline={(current bounding box.center)}]
\coordinate (in1) at (-1,0);
\coordinate (out1) at (1,0);
\coordinate (in2) at (-1,-1);
\coordinate (out2) at (1,-1);
\coordinate (x1) at (-0.35,0);
\coordinate (x2) at (-0.35,-1);
\coordinate (y1) at (0.35,0);
\coordinate (y2) at (0.35,-1);
\draw [dotted] (in1) -- (x1);
\draw [zUndirected] (x1) -- (y1);
\draw [dotted] (y1) -- (out1);
\draw [dotted] (in2) -- (x2);
\draw [dotted] (x2) -- (y2);
\draw [dotted] (y2) -- (out2);
\draw [photon] (x1) -- (x2);
\draw [photon] (y1) -- (y2);
\draw [fill] (x1) circle (.06);
\draw [fill] (x2) circle (.06);
\draw [fill] (y1) circle (.06);
\draw [fill] (y2) circle (.06);
\end{tikzpicture} 
+
\begin{tikzpicture}[baseline={(current bounding box.center)}]
\coordinate (in1) at (-1,0);
\coordinate (out1) at (1,0);
\coordinate (in2) at (-1,-1);
\coordinate (out2) at (1,-1);
\coordinate (x1) at (-0.35,0);
\coordinate (x2) at (-0.35,-1);
\coordinate (y1) at (0.35,0);
\coordinate (y2) at (0.35,-1);
\draw [dotted] (in1) -- (x1);
\draw [dotted] (x1) -- (y1);
\draw [dotted] (y1) -- (out1);
\draw [dotted] (in2) -- (x2);
\draw [zUndirected] (x2) -- (y2);
\draw [dotted] (y2) -- (out2);
\draw [photon] (x1) -- (x2);
\draw [photon] (y1) -- (y2);
\draw [fill] (x1) circle (.06);
\draw [fill] (x2) circle (.06);
\draw [fill] (y1) circle (.06);
\draw [fill] (y2) circle (.06);
\end{tikzpicture} \nonumber\\
=& \int\limits_q {e^{ib\cdot q}\, \dd(q\cdot u_1) \dd(q\cdot u_2) } ~ \left[ -{ ie^2 q_1 q_2 \gamma \over q^2} \right. \nonumber\\
 &  \left. + ie^4 q_1^2 q_2^2 { \big( (2D-7)\gamma^2 -1 \big) (m_1+m_2) \over 2 \big(\gamma^2-1\big) m_1 m_2}  G^{(1)}_i \right]\,,  
\end{align}
where we have adopted the notations in~\cite{Mogull:2020sak} for the integration measure and $\dd(x):=2\pi\delta(x)$, $\gamma = u_1\cdot u_2$ and $D$ denotes the spacetime dimension. The integral $G_i^{(1)}$ in $D=4-2\epsilon$ reads 
\begin{align}
G^{(1)}_i \!=\! \int\limits_{\ell_1}\!\! {\dd(\ell_1\cdot u_i) \over \ell_1^2 (q-\ell_1)^2 } \!=\!{(4\pi)^{\epsilon-\tfrac{3}{2}}\Gamma\left(\tfrac{1}{2}\!-\!\epsilon \right)^2  \Gamma\left(\tfrac{1}{2}\!+\!\epsilon \right)\over (-q^2)^{{1\over 2}+\epsilon} \Gamma(1-\epsilon)}\,.
\end{align}
Note that both the impact parameter $b^\mu$ and the total momentum transfer $q^\mu$ are spacelike and the Fourier transform is performed in $(D-2)$ dimensions due to the two $\delta$-functions as follows,
\begin{align}
\int\limits_{q} {e^{iq\cdot b} \over (-q^2)^\alpha} = { (-b^2)^{\alpha-1+\epsilon} \Gamma(1-\alpha-\epsilon) \over 4^\alpha \pi^{1-\epsilon} \sqrt{\gamma^2-1} \Gamma(\alpha) } \,.
\end{align}
Fourier transforming to the impact parameter space, we obtain
\begin{align}
\delta^{(0)} & = \alpha q_1 q_2 {\gamma\over \sqrt{\gamma^2-1}} {\Gamma(-\epsilon) \over \pi^{-\epsilon} } (\mathsf{b}^2)^\epsilon \,, \label{eq:phase1PM}\\
\delta^{(1)} & = - (\alpha q_1 q_2)^2 {\pi (m_1+m_2) \over 2 m_1 m_2 \sqrt{\gamma^2-1} \,\mathsf{b} } \,, \label{eq:phase2PM}
\end{align}
where $\mathsf{b} = |b| = \sqrt{-b^2}$ and $\alpha=e^2/(4\pi)$.\footnote{We multiply a factor of ${i\over (4\pi)^2} (4\pi e^{-\gamma_E})^\epsilon$ per loop in the end to restore the proper normalization~\cite{Jakobsen:2022fcj,DiVecchia:2021bdo}.}

Moving on to the next-to-next-to-leading order (3PM), we consider the comparable-masses sector ($m_1\sim m_2$) and the probe-limit sector ($m_1\ll m_2$ or $m_1\gg m_2$). The former contributes to both the conservative and the radiative parts of the scattering angle whereas the latter contributes only to the conservative part. In addition, the scattering angle begins to receive the so-called \emph{radiation reactions} at 3PM~\cite{Damour:2020tta,DiVecchia:2020ymx,DiVecchia:2021ndb,Herrmann:2021tct,Bjerrum-Bohr:2021din,Bern:2021xze,Saketh:2021sri} and we shall consider them separately.

\paragraph*{3PM Comparable Masses}
The conservative contribution from this sector is computed by the following diagrams,
\begin{align}\label{eq:RAm1m2}
&\left. i\delta^{(2)} \right|_{m_1 m_2} = 
\; {1\over 2}\left(
\begin{tikzpicture}[baseline={(current bounding box.center)}]
\coordinate (in1) at (-1,0);
\coordinate (out1) at (1,0);
\coordinate (in2) at (-1,-1);
\coordinate (out2) at (1,-1);
\coordinate (x1) at (-0.6,0);
\coordinate (y1) at (0,0);
\coordinate (z1) at (0.6,0);
\coordinate (x2) at (-0.6,-1);
\coordinate (y2) at (0,-1);
\coordinate (z2) at (0.6,-1);
\draw [dotted] (in1) -- (x1);
\draw [zUndirected] (x1) -- (y1);
\draw [dotted] (y1) -- (z1);
\draw [dotted] (z1) -- (out1);
\draw [dotted] (in2) -- (x2);
\draw [dotted] (x2) -- (y2);
\draw [zUndirected] (y2) -- (z2);
\draw [dotted] (z2) -- (out2);
\draw [photon] (x1) -- (x2);
\draw [photon] (y1) -- (y2);
\draw [photon] (z1) -- (z2);
\draw [fill] (x1) circle (.06);
\draw [fill] (y1) circle (.06);
\draw [fill] (z1) circle (.06);
\draw [fill] (x2) circle (.06);
\draw [fill] (y2) circle (.06);
\draw [fill] (z2) circle (.06);
\end{tikzpicture} +
\begin{tikzpicture}[baseline={(current bounding box.center)}]
\coordinate (in1) at (-1,0);
\coordinate (out1) at (1,0);
\coordinate (in2) at (-1,-1);
\coordinate (out2) at (1,-1);
\coordinate (x1) at (-0.6,0);
\coordinate (y1) at (0,0);
\coordinate (z1) at (0.6,0);
\coordinate (x2) at (-0.6,-1);
\coordinate (y2) at (0,-1);
\coordinate (z2) at (0.6,-1);
\draw [dotted] (in1) -- (x1);
\draw [dotted] (x1) -- (y1);
\draw [zUndirected] (y1) -- (z1);
\draw [dotted] (z1) -- (out1);
\draw [dotted] (in2) -- (x2);
\draw [zUndirected] (x2) -- (y2);
\draw [dotted] (y2) -- (z2);
\draw [dotted] (z2) -- (out2);
\draw [photon] (x1) -- (x2);
\draw [photon] (y1) -- (y2);
\draw [photon] (z1) -- (z2);
\draw [fill] (x1) circle (.06);
\draw [fill] (y1) circle (.06);
\draw [fill] (z1) circle (.06);
\draw [fill] (x2) circle (.06);
\draw [fill] (y2) circle (.06);
\draw [fill] (z2) circle (.06);
\end{tikzpicture} \right)
  \nonumber\\
&=  {ie^6 q_1^2 q_2^3\over 2 m_1 m_2}\int\limits_q e^{ib\cdot q}\prod_{i=1}^2 \dd(q\cdot u_i)\int\limits_{\ell_1\ell_2} {\dd(\ell_1\cdot u_2) \dd(\ell_2\cdot u_1) \over \ell_1^2 \ell_2^2 (q-\ell_1-\ell_2)^2} \nonumber\\ 
&\left[ - {\gamma (q-\ell_2)^2 \over (\ell_1\cdot u_1)^2} 
- {\gamma (q-\ell_1)^2 \over (\ell_2\cdot u_2)^2} + { \gamma^3 (q-\ell_1)^2 (q-\ell_2)^2 \over 2 (\ell_1\cdot u_1)^2 (\ell_2\cdot u_2)^2}\right. \nonumber\\
&\; \left. - {\gamma^2 q^2 \over (\ell_1\cdot u_1) (\ell_2\cdot u_2) } - {2 (\ell_1\cdot u_1) \over (\ell_2\cdot u_2)} -{2 (\ell_2\cdot u_2) \over (\ell_1\cdot u_1)}   \right]\,,
\end{align} 
where we have removed the tadpole terms from the integrand.\footnote{Tadpole terms are those that do not have all three massless poles $\ell_1^2$, $\ell_2^2$ and $(q-\ell_1-\ell_2)^2$ simultaneously, which integrate to zero.} 
Here we note that the two diagrams in the first line agree with the ``zigzag'' diagrams of the QED counterpart of HEFT. \footnote{More detailed discussions on the connections between the WQFT and HEFT approaches are given in Appendix \ref{app:comparewithHEFT}.}.
Using \textsf{LiteRed}~\cite{Lee:2012cn},~(\ref{eq:RAm1m2}) is cast by \emph{Integration-by-Parts}~(IBP) relations in the basis of master integrals as follows,\footnote{In all three sectors, only the principal-value parts of the time-symmetric propagators contribute to the real part of the phase $\delta$.Take the zigzag diagrams for example. The master integral $G_{1,1,0,0,1,1,1}$ needs to be dealt with in four sectors $G^{\pm \pm}_{1,1,0,0,1,1,1} $ seperately, depending on their respective $i\epsilon$-prescriptions for the two worldline propagators. However, rewriting $1/(x\pm i\epsilon) = \text{pv}(1/x) \mp i\pi \delta(x)$, the imaginary part drops out.}
\begin{align}\label{eq:RAm1m2afterIBP}
&\left. i\delta^{(2)} \right|_{m_1 m_2} = {ie^6 q_1^2 q_2^3\over 2m_1 m_2} \int\limits_q e^{ib\cdot q} \prod_{i=2}^2 \dd(q\cdot u_i) \Big[ a_1 G_{0,0,0,0,1,1,1} \nonumber\\
& \!\!+\! a_2 G_{0,0,0,0,1,2,1} \!+\! a_3 G_{0,0,0,0,2,1,1}\!+\!a_4 G_{1,1,0,0,1,1,1}\Big]\,.
\end{align}
The integrals are defined as 
\begin{align}\label{def:G}
G_{n_1 n_2\cdots n_7} = \int\limits_{\ell_1\ell_2}  {\dd(\ell_1\cdot u_2) \dd(\ell_2\cdot u_1) \over \rho_1^{n_1} \rho_2^{n_2} \cdots \rho_7^{n_7}}\,,
\end{align}
where the propagators are
\begin{align}
&\rho_1 = \ell_1\cdot u_1\,, ~~~ \rho_2 = \ell_2\cdot u_2\,, ~~~\rho_3 = \ell_1^2\,, ~~~ \rho_4 = \ell_2^2\,,\\
 &\rho_5 = (q-\ell_1-\ell_2)^2\,, ~~~\rho_6=(q-\ell_1)^2\,,~~~ \rho_7 = (q-\ell_2)^2\,. \nonumber
\end{align}
We list the coefficients in $D=4-2\epsilon$ below:
\begin{align}
a_1 & \!=\!-{2\epsilon \big(\gamma^4 (4 \epsilon^2 \!-\!2\epsilon \!-\!1)+\gamma^2 (2\epsilon \!+\! 3) -3\big) \over \gamma (\gamma^2-1) (1-2\epsilon)}\,,\\
a_2 & \!=\!{ (2\epsilon+1) \big( \gamma^4 (2\epsilon(6\epsilon \!-\!5)\!-\!1)+\gamma^2 (6\epsilon\!+\!3)\!-\!3\big) q^2 \over 3\gamma (\gamma^2-1)^2 (1-2\epsilon) \epsilon} \,,\\
a_3 & \!=\!  -{  \big( \gamma^4 (2\epsilon(6\epsilon \!-\!1)\!-\!5)+\gamma^2 (6\epsilon\!-\!3)\!+\!3\big) q^2 \over 3\gamma (\gamma^2-1) (1-2\epsilon)} \,,\\
a_4 & \!=\! -{\gamma^2 (2\gamma^2 \epsilon+3)\, q^2 \over 3 (\gamma^2-1)}\,.
\end{align}
These integrals are extensively studied in literature~\cite{Parra-Martinez:2020dzs,Smirnov:2012gma,Herrmann:2021tct,Brandhuber:2021eyq,Jakobsen:2022fcj} using differential equations~\cite{Kotikov:1990kg,Gehrmann:1999as,Lee:2014ioa,Henn:2013pwa,Henn:2014qga}. We display the explicit expressions for the \emph{real} parts of the master integrals present in (\ref{eq:RAm1m2afterIBP}) in Appendix \ref{app:masterintegrals} and (\ref{eq:RAm1m2afterIBP}) is readily evaluated. Fourier transforming to the impact parameter space and taking $\epsilon\rightarrow 0$, we obtain
\begin{align}\label{eq:ReRAm1m2}
 \left. \text{Re}\,\delta^{(2)} \right|_{m_1 m_2} =&  -{2 (\alpha q_1 q_2)^3  \,(\gamma^4-3\gamma^2+3) \over 3 m_1 m_2\,\mathsf{b}^2\, (\gamma^2-1)^{5/2} } \\
& + {2 (\alpha q_1q_2)^3  \gamma^2\big( \gamma \sqrt{ \gamma^2-1} - \text{arccosh}\gamma\big) \over  m_1 m_2\,\mathsf{b}^2 \,(\gamma^2-1)^{5/2} }  \,.\nonumber
\end{align}
The two terms are identified with the conservative and radiative contributions, because they result from boundary values computed in different regions. The first term comes purely from the potential region identified in~\cite{Brandhuber:2021eyq} while the second from the radiative region.
As will be demonstrated shortly, they reproduce the conservative and radiative parts of the scattering angle respectively. 

\paragraph*{3PM Probe Limit}
Similarly in the probe limit we consider the following diagrams,
\begin{align}\label{eq:RAprobe}
&\left. i\delta^{(2)} \right|_{m_2^2} \\
&= {1\over 3}\left(
\begin{tikzpicture}[baseline=-15pt]
\coordinate (in1) at (-1,0);
\coordinate (out1) at (1,0);
\coordinate (in2) at (-1,-1);
\coordinate (out2) at (1,-1);
\coordinate (x1) at (-0.6,0);
\coordinate (y1) at (0,0);
\coordinate (z1) at (0.6,0);
\coordinate (x2) at (-0.6,-1);
\coordinate (y2) at (0,-1);
\coordinate (z2) at (0.6,-1);
\draw [dotted] (in1) -- (x1);
\draw [zUndirected] (x1) -- (y1);
\draw [zUndirected] (y1) -- (z1);
\draw [dotted] (z1) -- (out1);
\draw [dotted] (in2) -- (x2);
\draw [dotted] (x2) -- (y2);
\draw [dotted] (y2) -- (z2);
\draw [dotted] (z2) -- (out2);
\draw [photon] (x1) -- (x2);
\draw [photon] (y1) -- (y2);
\draw [photon] (z1) -- (z2);
\draw [fill] (x1) circle (.06);
\draw [fill] (y1) circle (.06);
\draw [fill] (z1) circle (.06);
\draw [fill] (x2) circle (.06);
\draw [fill] (y2) circle (.06);
\draw [fill] (z2) circle (.06);
\end{tikzpicture} +
\begin{tikzpicture}[baseline=-15pt]
\coordinate (in1) at (-1,0);
\coordinate (out1) at (1,0);
\coordinate (in2) at (-1,-1);
\coordinate (out2) at (1,-1);
\coordinate (x1) at (-0.6,0);
\coordinate (y1) at (0,0);
\coordinate (z1) at (0.6,0);
\coordinate (x2) at (-0.6,-1);
\coordinate (y2) at (0,-1);
\coordinate (z2) at (0.6,-1);
\draw [dotted] (in1) -- (x1);
\draw [zUndirected] (x1) -- (y1);
\draw [zUndirected] (x1) ..controls (0,0.3).. (z1);
\draw [dotted] (y1) -- (z1);
\draw [dotted] (z1) -- (out1);
\draw [dotted] (in2) -- (x2);
\draw [dotted] (x2) -- (y2);
\draw [dotted] (y2) -- (z2);
\draw [dotted] (z2) -- (out2);
\draw [photon] (x1) -- (x2);
\draw [photon] (y1) -- (y2);
\draw [photon] (z1) -- (z2);
\draw [fill] (x1) circle (.06);
\draw [fill] (y1) circle (.06);
\draw [fill] (z1) circle (.06);
\draw [fill] (x2) circle (.06);
\draw [fill] (y2) circle (.06);
\draw [fill] (z2) circle (.06);
\end{tikzpicture} +
\begin{tikzpicture}[baseline=-15pt]
\coordinate (in1) at (-1,0);
\coordinate (out1) at (1,0);
\coordinate (in2) at (-1,-1);
\coordinate (out2) at (1,-1);
\coordinate (x1) at (-0.6,0);
\coordinate (y1) at (0,0);
\coordinate (z1) at (0.6,0);
\coordinate (x2) at (-0.6,-1);
\coordinate (y2) at (0,-1);
\coordinate (z2) at (0.6,-1);
\draw [dotted] (in1) -- (x1);
\draw [dotted] (x1) -- (y1);
\draw [zUndirected] (x1) ..controls (0,0.3).. (z1);
\draw [zUndirected] (y1) -- (z1);
\draw [dotted] (z1) -- (out1);
\draw [dotted] (in2) -- (x2);
\draw [dotted] (x2) -- (y2);
\draw [dotted] (y2) -- (z2);
\draw [dotted] (z2) -- (out2);
\draw [photon] (x1) -- (x2);
\draw [photon] (y1) -- (y2);
\draw [photon] (z1) -- (z2);
\draw [fill] (x1) circle (.06);
\draw [fill] (y1) circle (.06);
\draw [fill] (z1) circle (.06);
\draw [fill] (x2) circle (.06);
\draw [fill] (y2) circle (.06);
\draw [fill] (z2) circle (.06);
\end{tikzpicture} \right) \nonumber\\
& = {ie^6 q_1^3 q_2^3\over 12 m_1^2} \int\limits_q e^{ib\cdot q} \prod_{i=1}^2 \delta(q\cdot u_i) \int\limits_{\ell_1\ell_2} {\delta(\ell_1\cdot u_2) \delta(\ell_2\cdot u_2) \over \ell_1^2 \ell_2^2 (q-\ell_1-\ell_2)^2} \nonumber\\
&\!\!\left[ {\gamma \big( (q \!-\!\ell_1)^2\!-\!q^2\big) \over (\ell_1\cdot u_1)^2} \! +\! {2\gamma \big( (q\!-\!\ell_2)^2 \!-\!q^2\big) \over (\ell_2\cdot u_1)^2} \!+\! {\gamma^3\, q^2 (q\!-\!\ell_2)^2 \over (\ell_1\cdot u_1)^2 (\ell_2\cdot u_1)^2}\right. \nonumber\\
& \!\!\left. \!-\!{ \gamma^3 (q \!-\! \ell_2)^4 \over (\ell_1\cdot u_1)^2 (\ell_2\cdot u_1)^2} \!+\! { \gamma^3 (q\!-\!\ell_2)^2 \big( (q\!-\!\ell_2)^2 \!+\! (q\!-\!\ell_1)^2 \!-\!q^2\big) \over (\ell_1\cdot u_1)^3 (\ell_2\cdot u_1)} \right]\,.\nonumber
\end{align}
Here we have symmetrized the diagrams by labelling the momenta universally in all three diagrams. This symmetrization helps to reproduce all the pole structures expected explicitly in the classical limit of the corresponding Feynman diagrams in this sector. 
The two probe-limit sectors are simply related by relabelling $m_1\leftrightarrow m_2$.

After IBP reduction using \textsf{LiteRed}, the integrand in (\ref{eq:RAprobe}) is simplified to one single master integral
\begin{align}
{ie^6 q_1^3 q_2^3 (6\epsilon\!-\!1) \,\gamma\big( \gamma^2(6\epsilon\!-\!2)+3 \big) \over  6 m_1^2(\gamma^2-1)^2} G^{(2)}_2\,,
\end{align}
where the master integral in $D=4-2\epsilon$ reads 
\begin{align}
G^{(2)}_i & = \int\limits_{\ell_1\ell_2} {\dd(\ell_1\cdot u_i) \dd (\ell_2\cdot u_i) \over \ell_1^2 \ell_2^2 (q-\ell_1-\ell_2)^2}\nonumber\\
&= -{(4\pi^2)^{-3+2\epsilon} \over (-q^2)^{2\epsilon}} {\Gamma\left(\frac{1}{2}-\epsilon\right)^3 \Gamma(2\epsilon) \over \Gamma\left(\frac{3}{2}-3\epsilon\right)} \,.
\end{align}
That only one master integral contributes is also observed in the context of gravity~\cite{Brandhuber:2021eyq,Bjerrum-Bohr:2021wwt}. We note that the matching with the heavy-mass limit of scalar-QED in the probe limit is less manifest. The two integrands can be shown to be equal after IBP reduction. Fourier transforming to impact parameter space, we have
\begin{align}\label{eq:ReRAprobe}
\left. \text{Re}\,\delta^{(2)} \right|_{m_2^2} & = {(\alpha q_1 q_2)^3 \,\gamma(2\gamma^2-3) \over m_1^2 \,\mathsf{b}^2 \, (\gamma^2-1)^{5/2}}\,.
\end{align}
We will see shortly this reproduces the conservative part of the scattering angle in the probe limit.

\paragraph*{3PM Radiation Reaction}
The radiation reaction is accounted for by the following diagrams,
\begin{align}
&\left. i \delta^{(2)} \right|_{\text{r.r.}} = \left(
\begin{tikzpicture}[baseline={(current bounding box.center)}]
\coordinate (in1) at (-1,0);
\coordinate (out1) at (1,0);
\coordinate (in2) at (-1,-1);
\coordinate (out2) at (1,-1);
\coordinate (x1) at (-0.6,0);
\coordinate (y1) at (-0.25,0);
\coordinate (z1) at (0.25,0);
\coordinate (w1) at (0.6,0);
\coordinate (x2) at (-0.6,-1);
\coordinate (w2) at (0.6,-1);
\draw [dotted] (in1) -- (x1);
\draw [zUndirected] (x1) -- (y1);
\draw [dotted] (y1) -- (z1);
\draw [zUndirected] (z1) -- (w1);
\draw [dotted] (w1) -- (out1);
\draw [dotted] (in2) -- (x2);
\draw [dotted] (x2) -- (w2);
\draw [dotted] (w2) -- (out2);
\draw [photon] (x1) -- (x2);
\draw [photon] (w1) -- (w2);
\draw [photon] (y1) ..controls (0,-0.5).. (z1);
\draw [fill] (x1) circle (.06);
\draw [fill] (y1) circle (.06);
\draw [fill] (z1) circle (.06);
\draw [fill] (w1) circle (.06);
\draw [fill] (x2) circle (.06);
\draw [fill] (w2) circle (.06);
\end{tikzpicture}
+
\begin{tikzpicture}[baseline={(current bounding box.center)}]
\coordinate (in1) at (-1,-1);
\coordinate (out1) at (1,-1);
\coordinate (in2) at (-1,0);
\coordinate (out2) at (1,0);
\coordinate (x1) at (-0.6,-1);
\coordinate (y1) at (-0.25,-1);
\coordinate (z1) at (0.25,-1);
\coordinate (w1) at (0.6,-1);
\coordinate (x2) at (-0.6,0);
\coordinate (w2) at (0.6,0);
\draw [dotted] (in1) -- (x1);
\draw [zUndirected] (x1) -- (y1);
\draw [dotted] (y1) -- (z1);
\draw [zUndirected] (z1) -- (w1);
\draw [dotted] (w1) -- (out1);
\draw [dotted] (in2) -- (x2);
\draw [dotted] (x2) -- (w2);
\draw [dotted] (w2) -- (out2);
\draw [photon] (x1) -- (x2);
\draw [photon] (w1) -- (w2);
\draw [photon] (y1) ..controls (0,-0.5).. (z1);
\draw [fill] (x1) circle (.06);
\draw [fill] (y1) circle (.06);
\draw [fill] (z1) circle (.06);
\draw [fill] (w1) circle (.06);
\draw [fill] (x2) circle (.06);
\draw [fill] (w2) circle (.06);
\end{tikzpicture} \right) \nonumber \\
\!\!\!\!=& {ie^6 q_1^4 q_2^2 \over 2m_1^2}  \int\limits_q e^{ib\cdot q} \prod_{i=1}^2 \dd(q\cdot u_i) \int\limits_{\ell_1\ell_2} {\dd(\ell_1\cdot u_2) \dd(\ell_2 \cdot u_1) \over \ell_1^2 (q-\ell_1)^2 (q\!-\!\ell_1\!-\!\ell_2)^2} \nonumber\\
& \left[ 1 +{ (\ell_2\cdot u_2)^2 \over (\ell_1\cdot u_1)^2 }- {\gamma^2 q^2 \over 2(\ell_1\cdot u_1)^2} + {\gamma \ell_2^2 (\ell_2\cdot u_2) \over (\ell_1\cdot u_1)^3} \right. \nonumber\\
& \quad \left.+ {\gamma^2\ell_2^2 (q\!-\!\ell_2)^2 \over 2(\ell_1\cdot u_1)^4}\right]+ \big(\{q_1, m_1\}\leftrightarrow \{q_2,m_2\} \big) \,.
\end{align}
We again apply IBP reductions to the expression above, which leads to one single master integral $G_{0,0,1,0,1,1,0}$ as defined in (\ref{def:G}). Hence the radiation reaction contribution reads
\begin{align}\label{eq:RArr}
\!\!\!\!\left. \text{Re}\, \delta^{(2)} \right|_{\text{r.r.}}  = {-2 (\alpha q_1 q_2)^3\gamma^2 \over 3m_1 m_2\, \mathsf{b}^2 \,(\gamma^2-1)}\!\!\left[ {q_1/m_1 \over q_2/m_2} \!+\! {q_2/m_2 \over q_1/m_1}\right]\,.
\end{align}

\paragraph*{Scattering Angle}\label{sec:ScatteringAngle}
It is straightforward to compute the scattering angle in the center of mass frame via
\begin{align}
&\chi =- {\partial \delta \over \partial J}\,,
\end{align}
where $J$ denotes the total angular momentum and we have $J=\mathsf{p} \mathsf{b}$ with 
\begin{align}
\!\!\!\!\mathsf{p} = {m_1 m_2 \sqrt{\gamma^2-1} \over E}\,,~~
E  = \sqrt{m_1^2 + m_2^2 +2m_1 m_2\gamma}\,.
\end{align}
Plugging in (\ref{eq:phase1PM}) and (\ref{eq:phase2PM}), we obtain the scattering angle at 1PM and 2PM
\begin{align}
\chi^{(0)} &=   {2 \alpha q_1 q_2 E\,\gamma \over m_1 m_2 \, \mathsf{b} \, (\gamma^2-1)}\,,\\
\chi^{(1)} &= - {(\alpha q_1 q_2)^2 \pi \,E (m_1+m_2) \over 2 m_1^2 m_2^2 \,\mathsf{b}^2 \, (\gamma^2-1)}\,.
\end{align}

The conservative part of the 3PM scattering angle follows from the first term of the comparable-mass (\ref{eq:ReRAm1m2}) and the two probe-limit sectors~(\ref{eq:ReRAprobe}),
\begin{align}\label{eq:chiCon}
\chi^{(2)}_{\text{con}} =&-{4 (\alpha q_1 q_2)^3 E\,(\gamma^4 -3\gamma^2+3) \over 3 m_1^2 m_2^2 \,\mathsf{b}^3 \, (\gamma^2-1)^3}  \nonumber\\
& + {2 (\alpha q_1 q_2)^3 E (m_1^2+m_2^2) \,\gamma(2\gamma^2-3) \over 3 m_1^3 m_2^3 \,\mathsf{b}^3\, (\gamma^2-1)^3} \,.
\end{align}
Likewise, the radiative part at 3PM follows from the second line of (\ref{eq:ReRAm1m2}) and the radiation reaction term (\ref{eq:RArr}),
\begin{align}\label{eq:chiRad}
\chi^{(2)}_{\text{rad}} = & \quad {4 (\alpha q_1 q_2)^3 E\,\gamma^2 \over m_1^2 m_2^2 \, \mathsf{b}^3 } \left( {\gamma \over (\gamma^2-1)^{5/2}} - {\text{arccosh}\gamma \over (\gamma^2-1)^3}  \right) \nonumber\\
& - {4 (\alpha q_1 q_2)^3 E\,\gamma^2 \over 3 m_1^2 m_2^2 (\gamma^2-1)^{3/2}} \left( {q_1/m_1 \over q_2/m_2} + {q_2/m_2 \over q_1/m_1}\right)\,.
\end{align}
For (\ref{eq:chiCon}) and (\ref{eq:chiRad}) we find agreement with known results in literature~\cite{Bern:2021xze,Saketh:2021sri}.

\paragraph*{Discussions}
We have demonstrated a highly streamlined method for obtaining both the conservative and radiative contributions of the scattering angle in the WQFT formalism for scalar-QED. The scattering angle is computed from a generating function that naturally arises from the WQFT path integral. This generating function is constructed to be purely classical by virtue of WQFT and coincides with the recently proposed ``HEFT phase'', although the precise connection between the two remains to be clarified. Its real part also agrees with the radial action, while the differences between their respective imaginary parts are yet to be investigated. These observations are expected to hold in other WQFTs, especially those in a gravitational background, which we leave to future work. It is also interesting to further clarify the relation between this generating function and the eikonal phase in the context of WQFT. Another immediate followup is to study higher PM orders. In particular, the probe limit involves only one diagram (up to symmetrization) at any order, for which the vertices are known on closed forms. In this limit, it is promising to obtain all-loop results from WQFT.

\paragraph*{Acknowledgements}
TW is grateful to G.~Mogull and J.~Plefka for initial collaborations, many discussions and useful comments on the manuscript. TW thanks R.~Bonezzi, M.~V.~S.~Saketh, C.~Heissenberg, G.~Chen, A.~Brandhuber and G.~Travaglini for discussions and clarifications on their respective works. The research is supported by the Special Research Assistantship of the Chinese Academy of Sciences and Humboldt Universität zu Berlin. It is also supported in part by the Key Research Program of the Chinese Academy of Sciences, Grant NO. XDPB15, and by National Natural Science Foundation of China under Grant No.11935013,~11947301,~12047502,~12047503. TW is also supported by the Fellowship of China Postdoctoral Science Foundation (No. 2022M713228).

\appendix
\section{WQFT Feynman Rules for Sacalar-QED}\label{app:FeynRules}
The WQFT Feynman rules can be read off from the action with the trajectory $x^\mu(\sigma) = b^\mu + u^\mu \sigma+z^\mu(\sigma)$ and the plane wave $A_\mu (x) = \sum_{i=1}^n \varepsilon_{i\mu} e^{ik_i\cdot x}$ plugged in. The interaction term of the worldline coupled to one photon then reads 
\begin{align}
\mathcal S_{\text{int}} =& \sum_{n=0}^\infty \frac{i^n e\, q_a}{n!} \int\limits_{k,\omega_1,\cdots,\omega_n} e^{ik\cdot b} \dd\left(k\cdot u+\sum_i\omega_i\right) \prod_i z^{\rho_i}\nonumber\\
&  A_\mu(-k)\left( \prod_i k_{\rho_i} u^\mu+\sum_{i}\omega_i \delta^\mu_{\rho_i}\prod_{j\neq i}k_{\rho_j} \right)\,.
\end{align}
Hence the momentum-space Feynman rule for the worldline-photon interaction at the zeroth order of $z^\rho$ is given by 
\begin{align}\label{eq:vertex0}
\begin{tikzpicture}[baseline={(current bounding box.center)}]
\coordinate (in) at (-1,0);
\coordinate (out) at (1,0);
\coordinate (x) at (0,0);
\node (k) at (0,-1.3) {$A_\mu(k)$};
\draw [dotted] (in) -- (x);
\draw [dotted] (x) -- (out);
\draw [graviton] (x) -- (k);
\draw [fill] (x) circle (.08);
\end{tikzpicture} =
-ie\,q_a\text{exp}\left(ik\cdot b\right)\dd(k\cdot u) u^\mu \,.
\end{align}
At the linear order of $z^\rho$, we have
\begin{align}
\!\!\!\!\!\!\!\!
\begin{tikzpicture}[baseline={(current bounding box.center)}]
\coordinate (in) at (-0.75,0);
\coordinate (out) at (0.75,0);
\coordinate (x) at (0,0);
\node (k) at (0,-1.3) {$A_\mu(k)$};
\draw (out) node [right] {$z^\rho(\omega)$};
\draw [dotted] (in) -- (x);
\draw [zUndirected] (x) -- (out);
\draw [graviton] (x) -- (k);
\draw [fill] (x) circle (.075);
\end{tikzpicture} \!=\! e\, q_a e^{ik\cdot b}\dd(k\cdot u\!+\!\omega) \left( k_\rho u^\mu \!+\! \omega \delta^\mu_\rho\right)\,.
\end{align}
Finally, at the quadratic order, the worldline vertex reads
\begin{align}
 \begin{tikzpicture}[baseline=-20pt]
\coordinate (in) at (-1,0);
\coordinate (out1) at (1,0);
\coordinate (out2) at (1,0.5);
\coordinate (x) at (0,0);
\coordinate (kk) at (0,-1);
\node (k) at (-0.5,-0.8) {$A_\mu$};
\draw (out1) node [right] {$z^{\rho_1}$};
\draw (out2) node [right] {$z^{\rho_2}$};
\draw [dotted] (in) -- (x);
\draw [zUndirected] (x) -- (out1);
\draw [zUndirected] (x) to[out=40,in=180] (out2);
\draw [graviton] (x) -- (kk);
\draw [fill] (x) circle (.08);
\end{tikzpicture} =& i e\, q_a e^{ik\cdot b}\dd\left(k\cdot u+ \omega_1+\omega_2\right)\\ 
&\left(k_{\rho_1} k_{\rho_2} u^\mu + \omega_1 \delta^\mu_{\rho_1}k_{\rho_2}+ \omega_2 \delta^\mu_{\rho_2}k_{\rho_1}\right) \,. \nonumber
\end{align}
This completes all the worldline interactions needed for the computation in the main text. 

In general, for a worldline vertex emitting a photon with $n$ deflections, the Feynman rule is at the $n$-th order of $z^\rho$ and satisfies the same recursion relation as in~\cite{Mogull:2020sak}
\begin{align}
& {\partial \over \partial b^\nu} V_{\rho_1\cdots \rho_n}^\mu(k,\omega_1,\cdots,\omega_n)  \nonumber\\
=& \left.V^\mu_{\rho_1,\cdots,\rho_{n+1}} (k,\omega_1,\cdots,\omega_n,0) \right|_{\rho_{n+1} = \nu} \,.
\end{align}

\section{Master Integrals}\label{app:masterintegrals}
Here we list the explicit expressions for the \emph{real} part of the master integrals in $D=4-2\epsilon$ used in the main text, which can be found in~\cite{Parra-Martinez:2020dzs,Brandhuber:2021eyq,Jakobsen:2022fcj}. For detailed discussions on these integrals including the imaginary part, see~\cite{Damgaard:2021ipf,Brandhuber:2021eyq,Bjerrum-Bohr:2021wwt}.   For convenience, we adopt the basis below
\begin{align}
I_1 &= (-q^2)^{2\epsilon} G_{0,0,0,0,1,1,1}\,,\\
I_2 &= (-q^2)^{1+2\epsilon} G_{0,0,0,0,2,1,1}\,,\\
I_3 &= (-q^2)^{1+2\epsilon} G_{0,0,0,0,1,2,1}\,,\\
I_4 &= (-q^2)^{1+2\epsilon} G_{1,1,0,0,1,1,1}\,,\\
I_5 &= (-q^2)^{2\epsilon} G_{0,0,1,0,1,1,0}\,.
\end{align}
Then the real parts of the integrals above read
\begin{align}
I_1  =& -{(4\pi^2)^{-3+2\epsilon}\,\text{arccosh}\gamma \over \sqrt{\gamma^2-1}\epsilon}\,,\\
I_2  =& {\color{gray} -{(4\pi^2)^{-3+2\epsilon} \over \sqrt{\gamma^2-1} \,\epsilon}} \,, \\
I_3  =& -{(4\pi^2)^{-3+2\epsilon} \over 1+2\epsilon} \left( 6\sqrt{\gamma^2-1}\,\epsilon\,\text{arccosh}\gamma+{2\gamma \over \sqrt{\gamma^2-1}} \right) \nonumber\\
& {\color{gray} + {(4\pi^2)^{-3+2\epsilon} \sqrt{\gamma^2-1} \over 1+2\epsilon} }\,,\\
I_4 =& {\color{gray}{(4\pi^2)^{-3+2\epsilon} \,2\text{arccosh}\gamma\over (\gamma^2-1)\epsilon}}\,,\\
I_5 = & {\color{gray} {(4\pi^2)^{-3+2\epsilon} \sqrt{\gamma^2-1} \over 8\epsilon^2 -6\epsilon+1}} \,.
\end{align}
For each integral, the contribution from the potential region is colored black while that from the radiative region is colored gray. The conservative/radiative part computed in the main text can be obtained by restricting the integrals above to their respective potential/radiative regions. 

\section{Comparison With HEFT}\label{app:comparewithHEFT}
Here we demonstrate the matching between WQFT graphs and HEFT ones. On the WQFT side, consider a single photon emitted from a worldline and we have
\begin{align}
\begin{tikzpicture}[baseline={(current bounding box.center)}]
\coordinate (in) at (-1,0);
\coordinate (out) at (1,0);
\coordinate (x) at (0,0);
\node (k) at (0,-1.3) {$\varepsilon^\mu_1(\ell_1)$};
\draw [dotted] (in) -- (x);
\draw [dotted] (x) -- (out);
\draw [photon] (x) -- (k);
\draw [fill] (x) circle (.08);
\end{tikzpicture} \propto ie q_1(\varepsilon_1\cdot u_1) = {1\over m_1} A_3^{\text{sQED}}(\varepsilon_1,u_1)\,,
\end{align}
where we have omitted the factor $e^{i\ell_1\cdot b_1}$ and the $\delta$-function in the WQFT Feynman rule~(\ref{eq:vertex0}), which will be restored in order to assemble the relevant integrand. $A_3^{\text{sQED}}(2,u_1)$ denotes the 3-point HEFT amplitude in scalar-QED and we follow the notation in~\cite{Brandhuber:2021eyq}. Similarly, it is straightforward to check that the following WQFT diagram is proportional to the 4-point HEFT amplitude up to a trivial overall factor $e^{i(\ell_1-\ell_2)\cdot b_1}$ and the $\delta$-function in the WQFT Feynman rule:
\begin{align}
&\begin{tikzpicture}[baseline={(current bounding box.center)}]
\coordinate (in1) at (-1,0);
\coordinate (out1) at (1,0);
\coordinate (x1) at (-0.35,0);
\node (x2) at (-0.35,-1.2) {$\varepsilon_1$};
\coordinate (y1) at (0.35,0);
\node (y2) at (0.35,-1.2) {$\varepsilon_2$};
\draw [dotted] (in1) -- (x1);
\draw [zUndirected] (x1) -- (y1);
\draw [dotted] (y1) -- (out1);
\draw [photon] (x1) -- (x2);
\draw [photon] (y1) -- (y2);
\draw [fill] (x1) circle (.06);
\draw [fill] (y1) circle (.06);
\end{tikzpicture} \nonumber\\
\propto & \quad {ie^2 q_1^2 (\ell_1\cdot \ell_2) (\varepsilon_1\cdot u_1) (\varepsilon_2\cdot u_1) \over m_1 (\ell_1\cdot u_1)^2} +{ie^2 q_1^2 (\varepsilon_1\cdot \varepsilon_2)\over m_1}\nonumber\\
& + {ie^2 q_1^2 (\ell_1\cdot \varepsilon_2) (\varepsilon_1\cdot u_1) \over m_1(\ell_1\cdot u_1)}-{ie^2 q_1^2 (\ell_2\cdot \varepsilon_1) (\varepsilon_2\cdot u_1) \over m_1(\ell_1\cdot u_1)}\nonumber\\
&={1\over m_1} A_4^{\text{sQED}}(\varepsilon_1,\varepsilon_2,u_1)\,.
\end{align}
At 1PM and 2PM the WQFT graphs in Eq.~(4) in the main text are now readily identified with the HEFT integrand. For instance,
\begin{align}
&\begin{tikzpicture}[baseline={(current bounding box.center)}]
\coordinate (in1) at (-1,0);
\coordinate (out1) at (1,0);
\coordinate (in2) at (-1,-1);
\coordinate (out2) at (1,-1);
\coordinate (x1) at (-0.35,0);
\coordinate (x2) at (-0.35,-1);
\coordinate (y1) at (0.35,0);
\coordinate (y2) at (0.35,-1);
\draw [dotted] (in1) -- (x1);
\draw [zUndirected] (x1) -- (y1);
\draw [dotted] (y1) -- (out1);
\draw [dotted] (in2) -- (x2);
\draw [dotted] (x2) -- (y2);
\draw [dotted] (y2) -- (out2);
\draw [photon] (x1) -- (x2);
\draw [photon] (y1) -- (y2);
\draw [fill] (x1) circle (.06);
\draw [fill] (x2) circle (.06);
\draw [fill] (y1) circle (.06);
\draw [fill] (y2) circle (.06);
\end{tikzpicture} = {\color{gray}\int\limits_q {e^{ib\cdot q}\, \delta(q\cdot u_1) \delta(q\cdot u_2) }\int\limits_{\ell_1} {\delta\left(\ell_1\cdot u_2 \right) }}\nonumber\\
& \sum_{h_1,h_2} {A^{h_1,h_2}_4(\ell_1,\ell_2,u_1) A_3^{-h_1}(\ell_1,u_2) A_3^{-h_2}(\ell_2,u_2) \over m_1 {\ell_1^2 \ell_2^2}}\,,
\end{align}
where $\ell_1+\ell_2=q$ and $h_1,h_2$ denote the helicities of the photons. The first line in gray comes from the factors we have ignored above. The $\delta$-function is precisely the massive ``cut-propagator'' in the HEFT computation. At 3PM, a graph-to-graph matching in the same fashion can be easily seen in the comparable-masse sector. In the probe-limit sector, the integrands computed from WQFT and HEFT can be shown to be equivalent up to IBP relations.

\bibliography{main}
\end{document}